\documentclass[conference]{IEEEtran}
\usepackage{amsmath, amssymb}

\usepackage{psfrag}
\usepackage{graphicx}
\begin{document}
% paper title
\title{A Note on Multiple-Access Channels with Strictly-Causal State
  Information} 
\author{\IEEEauthorblockN{Amos Lapidoth}
\IEEEauthorblockA{ETH Zurich\\Switzerland\\
Email: lapidoth@isi.ee.ethz.ch}
\and
\IEEEauthorblockN{Yossef Steinberg}
\IEEEauthorblockA{Technion---Israel Institute of Technology\\Israel\\
Email: ysteinbe@ee.technion.ac.il}
}
\maketitle
\newcommand{\set}[1]{\mathcal{#1}} 
\newcommand{\Normal}[2]{\mathcal{N}\!\left({#1},{#2}\right)} %Normal
\newcommand{\E}[2][]{\textnormal{\textsf{E}}_{#1}\!\left[#2\right]} %expectatio
\newcommand{\Prv}[1]{\Pr[#1]} %usage: Pr[X \leq 5]
\newcommand{\bigPrv}[1]{\Pr\bigl[#1\bigr]} 
\newcommand{\BigEcond}[3][]{\textnormal{\textsf{E}}_{#1}\!\Bigl[#2
  \kern-0.1em\Bigm|\kern-0.1em #3\Bigr]}
\newcommand{\vect}[1]{\mathbf{#1}}
\newcommand{\bfy}{\bml{y}}
\newcommand{\bfw}{\bml{w}}
\newcommand{\eps}{\epsilon} %shorthand for epsilon
\newcommand{\markov}{\textnormal{\mbox{$\multimap\hspace{-0.73ex}-\hspace{-2ex}-$}}}
\providecommand{\abs}[1]{\lvert#1\rvert}
\providecommand{\bigabs}[1]{\bigl\lvert#1\bigr\rvert}

\newcommand{\bigIntRange}[2]{\bigl[#1 : #2\bigr]}
\newcommand{\IntRange}[2]{[#1 : #2]}

\newcommand{\Scom}{W} % The common state - random variable
\newcommand{\ScomAlph}{\mathcal{W}} % The common state alphabet
\newcommand{\scom}{w} % The common state - realization
\newcommand{\bfscom}{\bfw} % common state sequence

\newcommand{\CapSCcom}{\mathcal{C}_{\textnormal{s-c}}^{\textnormal{com}}}
\newcommand{\pmfSCcom}{\mathcal{P}_{\textnormal{s-c}}^{\textnormal{com}}}
\newcommand{\RSCcom}{\mathcal{R}_{\textnormal{s-c}}^{\textnormal{com}}}
\newcommand{\OutSCcomOne}{\mathcal{O}_{\textnormal{s-c},1}^{\textnormal{com}}}
\newcommand{\OutSCcomTwo}{\mathcal{O}_{\textnormal{s-c},2}^{\textnormal{com}}}
\newcommand{\OutSCcomThree}{\mathcal{O}_{\textnormal{s-c},3}^{\textnormal{com}}}

\newcommand{\CapCcom}{\mathcal{C}_{\textnormal{cau}}^{\textnormal{com}}}
\newcommand{\pmfCcom}{\mathcal{P}_{\textnormal{cau}}^{\textnormal{com}}}
\newcommand{\RCcom}{\mathcal{R}_{\textnormal{cau}}^{\textnormal{com}}}

\newcommand{\CapSCind}{\mathcal{C}_{\textnormal{s-c}}^{\textnormal{ind}}}
\newcommand{\pmfSCind}{\mathcal{P}_{\textnormal{s-c}}^{\textnormal{ind}}}
\newcommand{\RSCind}{\mathcal{R}_{\textnormal{s-c}}^{\textnormal{ind}}}

\newcommand{\CapCind}{\mathcal{C}_{\textnormal{cau}}^{\textnormal{ind}}}
\newcommand{\pmfCind}{\mathcal{P}_{\textnormal{cau}}^{\textnormal{ind}}}
\newcommand{\RCind}{\mathcal{R}_{\textnormal{cau}}^{\textnormal{ind}}}

\newtheorem{problem}{Problem}
\newtheorem{definition}{Definition}
\newtheorem{lemma}{Lemma}
\newtheorem{proposition}{Proposition}
\newtheorem{corollary}{Corollary}
\newtheorem{example}{Example}
\newtheorem{note}{Note}
\newtheorem{conjecture}{Conjecture}
\newtheorem{algorithm}{Algorithm}
\newtheorem{theorem}{Theorem}
\newtheorem{exercise}{Exercise}
\newtheorem{remark}{Remark}

\newcommand{\bml}[1]{\vect{#1}}
\newcommand{\ssbml}[1]{\scriptsize{\mbox{\boldmath $ #1 $}}}
\newcommand{\sbml}[1]{\mbox{\scriptsize{\boldmath $ #1 $}}}

\newcommand{\be}{\begin{equation}}
\newcommand{\ee}{\end{equation}}
\newcommand{\bea}{\begin{eqnarray}}
\newcommand{\eea}{\end{eqnarray}}
\newcommand{\beaa}{\begin{eqnarray*}}
\newcommand{\eeaa}{\end{eqnarray*}}
\newcommand{\ben}{\begin{enumerate}}
\newcommand{\een}{\end{enumerate}}
\newcommand{\bi}{\begin{itemize}}
\newcommand{\ei}{\end{itemize}}
\newcommand{\eqd}{\stackrel{\Delta}{=}}
\newcommand{\limn}{\lim_{n\rightarrow\infty}}

\newcommand{\one}{\frac{1}{n}}
\newcommand{\half}{\frac{1}{2}}
\newcommand{\onei}{{\rm 1\!\!\!\:I}}
\newcommand{\NN}{{\rm I\!\!\!\;N}}

\newcommand{\Prvcond}[2]{\Pr[#1 \kern0.1em|\kern0.1em #2]} %cond. prob.
\newcommand{\bigPrvcond}[2]{\Pr\bigl[#1 \kern-0.1em \bigm| \kern-0.1em#2\bigr]}
\newcommand{\BigPrvcond}[2]{\Pr\Bigl[#1 \kern-0.1em \Bigm| \kern-0.1em#2\Bigr]}
\newcommand{\biggPrvcond}[2]{\Pr\biggl[#1 \kern-0.1em \biggm| \kern-0.1em#2\biggr]}
\newcommand{\BigPrv}[1]{\Pr\Bigl[#1\Bigr]}

\begin{abstract}
  We propose a new inner bound on the capacity region of a memoryless
  multiple-access channel that is governed by a memoryless state that
  is known strictly causally to the encoders. The new inner bound
  contains the previous bounds, and we provide an example demonstrating
  that the inclusion can be strict. 

  A variation on this example is then applied to the case where the
  channel is governed by two independent state sequences, where each
  transmitter knows one of the states strictly causally. The example
  proves that, as conjectured by Li \emph{et al.}, an inner bound that they
  derived for this scenario can indeed by strictly better than previous
  bounds.
\end{abstract}

\section{Introduction}

If a memoryless single-user channel is governed by an independent and
identically distributed (IID) state sequence, then its capacity is not
increased if the state is made available to the encoder in a
strictly-causal way. The picture changes dramatically on the
multiple-access channel (MAC) \cite{LapidothSteinbergIZS10},
\cite{LapidothSteinbergISIT10}: In the ``single-state scenario,''
where the channel is governed by a single state sequence, the capacity
region typically increases if the state is revealed to both
transmitters in a strictly causal way
\cite{LapidothSteinbergIZS10}. Some of the gains can be attributed to
the ability of the two encoders to compress the state information and
to cooperate in sending the compressed version to the receiver. But
strictly-causal side information (SI) is beneficial even in the
``double-state scenario,'' where the channel is governed by two
independent states, with each transmitter knowing one of the sequences
strictly causally. In this case too, the side information can be
helpful even though the transmitters cannot cooperate in compressing
the states or in sending them \cite{LapidothSteinbergISIT10}.

The present note deals with both the single-state and the double-state
scenarios. For the single-state scenario, we present a new inner bound
on the capacity region. This bound contains the inner bound of
\cite{LapidothSteinbergIZS10} (which was extended to the
many-transmitters scenario in \cite{LiSimeoneYenerArxiv}). We also
provide an example showing that the inclusion can be strict. 

By adapting this example to the double-state scenario, we provide an
example showing that---as conjectured in
\cite{LiSimeoneYenerArxiv}---the inner bound proposed by Li \emph{et
  al.} in \cite{LiSimeoneYenerArxiv} can be strictly larger than that
in \cite{LapidothSteinbergISIT10}.

To keep the contribution focused, we do not consider causal side
information in this note, although our results can be carried over to
that setting as in \cite{LapidothSteinbergIZS10},
\cite{LapidothSteinbergISIT10}.

We next describe the two scenarios more explicitly. Our descriptions
are identical to those in \cite{LapidothSteinbergIZS10},
\cite{LapidothSteinbergISIT10} except that, for simplicity, we do not
consider cost constraints and we assume throughout that all the
alphabets are finite.

\subsection{The Single-State Scenario}

In the single-state scenario we are given a discrete memoryless
state-dependent MAC of law $P_{Y|\Scom,X_1,X_2}$ with state alphabet
$\ScomAlph$, state probability mass function (PMF) $P_{\Scom}$, input
alphabets ${\cal X}_1$ and ${\cal X}_2$, and output alphabet ${\cal
  Y}$. Sequences of letters from $\ScomAlph$ are denoted
$\scom^n=(\scom_{1},\scom_{2},\ldots,\scom_{n})$ and $\scom_{i}^{j}
=(\scom{i},\scom_{i+1}\ldots,\scom_{j})$. Similar notation holds for
all alphabets, e.g. $x_1^n=(x_{1,1},x_{1,2},\ldots,x_{1,n})$,
$x_{2,i}^j=(x_{2,i},x_{2,i+1},\ldots,x_{2,j})$. When there is no risk
of ambiguity, $n$-sequences will sometimes be denoted by boldface
letters, $\bfy$, $\bml{x}_1$, $\bfscom$, etc.  The laws governing
$n$-sequences of output letters and states are 
\begin{equation*}
P^n_{Y|\Scom,X_1,X_2}(\bfy|\bfscom,\bml{x}_1,\bml{x}_2)
= \prod_{i=1}^n P_{Y|\Scom,X_1,X_2}(y_i|\scom_{i},x_{1,i},x_{2,i}),
\end{equation*}
\[
P^n_{\Scom}(\bfscom) = \prod_{i=1}^n P_{\Scom}(\scom_{i}).
\]
For notational convenience, we henceforth omit the superscript~$n$,
and we denote the channel by $P$.
%Let $\phi_k\colon{\cal X}_k\rightarrow [0,\infty)$, $k=1,2$, be single letter cost functions.
%The cost associated with the transmission of sequence $\bml{x}_k$ at input $k$ is
%defined as
%\[
%\phi_k(\bml{x}_k) = \one\sum_{i=1}^n \phi_k(x_{k,i}), \quad k\in \{1,2\}.
%\]

\begin{definition}
\label{def:1}
Given positive integers $\nu_1$, $\nu_2$, let $\mathcal{M}_1$ denote
the set $\{1,2,\ldots,\nu_1\}$, and let $\mathcal{M}_2$ denote the set
$\{1,2,\ldots,\nu_2\}$.  An $(n,\nu_1,\nu_2,\epsilon)$ code with
strictly-causal side information (SI) at the encoders is a pair of
sequences of encoder mappings \be f_{k,i}\colon
\ScomAlph^{i-1}\times\mathcal{M}_k \rightarrow {\cal X}_k,\ \ k=1,2,\
\ i=1,\ldots,n
\label{eq:1}
\ee
and a decoding mapping
\[
g\colon {\cal Y}^n\rightarrow \mathcal{M}_1\times \mathcal{M}_2
\]
such that 
% the input costs are bounded by $\Gamma_k$
% \[
% \phi_k(\bml{x}_k)\leq \Gamma_k,\ \ \ k=1,2,
% \]
% and 
the average probability of error $P_{\text{e}}$ does now
exceed~$\eps$. Here $P_{\text{e}}$ is $1 - P_{\text{c}}$;
\begin{subequations}
  \begin{equation*}
    P_{\text{c}} = \frac{1}{\nu_1 \nu_2} 
 \sum_{m_1=1}^{\nu_1} \sum_{m_2=1}^{\nu_2} \Pr(\text{correct}|m_{1}, m_{2});
  \end{equation*}
  and
  \begin{multline*}
    \Pr(\text{correct}|m_{1}, m_{2}) = \\ \sum_{\bfscom}
    P_{\Scom}(\bfscom)
    P\left(g^{-1}(m_1,m_2)|\bfscom,\bml{f}_1(\bfscom,m_1),\bml{f}_2(\bfscom,m_2)\right),
  \end{multline*}
\end{subequations}
where $g^{-1}(m_1,m_2)\subset{\cal Y}^n$ is the decoding set of the
pair of messages $(m_1,m_2),$ and
\[
\bml{f}_k(\bfscom,m_k)=
(f_{k,1}(m_k),f_{k,2}(\scom_{1},m_k),\ldots,f_{k,n}(\scom^{n-1},m_k)).
\]
\end{definition}
The rate pair $(R_1,R_2)$ of the code is defined as
\[
R_1=\one\log \nu_1,\ \ \ \ R_2=\one\log \nu_2.
\]
A rate-pair $(R_1,R_2)$ is said to be achievable if for every positive
$\epsilon$ and sufficiently large $n$ there exists an
$(n,2^{nR_1},2^{nR_2},\epsilon)$ code with strictly-causal SI for the
channel $P_{Y|\Scom,X_1,X_2}$.  The capacity region of the channel
with strictly-causal SI is the closure of the set of all achievable
pairs $(R_1,R_2)$, and is denoted $\CapSCcom$. The subscript
``s-c'' stands for strictly-causal.  % Our interest is in $\CapSCcom$.

\subsection{The Double-State Scenario}

In the double-state scenario we are given a discrete memoryless
state-dependent MAC $P_{Y|S_1,S_2,X_1,X_2}$ with state alphabets
${\cal S}_1$ and ${\cal S}_2$, state probability mass functions (PMFs)
$P_{S_1}$ and $P_{S_2}$, input alphabets ${\cal X}_1$ and ${\cal
  X}_2$, and output alphabet ${\cal Y}$.
%We use notation similar to
%Section~\ref{sec:res}, except that the subscript $c$ of the common
%state is now replaced with the state index. Thus, sequences of letters
%from ${\cal S}_k$ are denoted by
%$s_k^n=(s_{k,1},s_{k,2},\ldots,s_{k,n})$ and
%$s_{k,i}^j=(s_{k,i},s_{k,i+1}\ldots,s_{k,j})$, $k=1,2$.  
The laws governing $n$ sequences of output letters and states are 
\begin{multline*}
P^n_{Y|S_1,S_2,X_1,X_2}(\bfy|\bml{s}_1\bml{s}_2,\bml{x}_1,\bml{x}_2) \\
=
\prod_{i=1}^n P_{Y|S_1,S_2,X_1,X_2}(y_i|s_{1,i},s_{2,i},x_{1,i},x_{2,i}),
\end{multline*}
\begin{equation*}
P^n_{S_1,S_2}(\bml{s}_1,\bml{s}_2) = \prod_{i=1}^n P_{S_1}(s_{1,i})P_{S_2}(s_{2,i}).
\end{equation*}
For notational convenience, we henceforth omit the superscript~$n$,
and we denote the channel by $P$.
%Let $\phi_k\colon{\cal X}_k\rightarrow [0,\infty)$, $k=1,2$, be single letter cost functions.
%The cost associated with the transmission of sequence $\bml{x}_k$ at input $k$ is
%defined as
%\[
%\phi_k(\bml{x}_k) = \one\sum_{i=1}^n \phi_k(x_{k,i}).
%\]

Given positive integers $\nu_1$, $\nu_2$, let $\mathcal{M}_1$
be the set $\{1,2,\ldots,\nu_1\}$
and $\mathcal{M}_2$ the set $\{1,2,\ldots,\nu_2\}$.
An $(n,\nu_1,\nu_2,\epsilon)$ code with strictly
causal independent SI at the encoders is a pair of sequences of encoder
mappings
\be
f_{k,i}\colon {\cal S}_k^{i-1}\times\mathcal{M}_k \rightarrow  {\cal X}_k,\ \ k=1,2,\ \ i=1,\ldots,n
\label{eq:P2_0}
\ee
and a decoding mapping
\[
g\colon {\cal Y}^n\rightarrow \mathcal{M}_1\times \mathcal{M}_2
\]
such that 
% the input costs are bounded by $\Gamma_k$
% \[
% \phi_k(\bml{x}_k)\leq \Gamma_k,\ \ \ k=1,2,
% \]
% and 
the average probability of error $P_e$ is bounded by $\epsilon$, where
$P_{\text{e}} = 1 - P_{\text{c}}$ and 
\begin{multline*}
P_\text{c} = \frac{1}{\nu_1\nu_2} \sum_{m_1=1}^{\nu_1}
\sum_{m_2=1}^{\nu_2}  \sum_{\bml{s}_1,\bml{s}_2} \\
P_{S_1}(\bml{s}_1) P_{S_2}(\bml{s}_2) 
P \! \left(g^{-1}(m_1,m_2)|\bml{s}_1,\bml{s}_2,
\bml{f}_1(\bml{s}_1,m_1),\bml{f}_2(\bml{s}_2,m_2)\right)
\end{multline*}
where $g^{-1}(m_1,m_2)\subset{\cal Y}^n$ is the decoding set of the pair of messages $(m_1,m_2),$ 
and
\[
\bml{f}_k(\bml{s}_k,m_k)=
\bigl( f_{k,1}(m_k),f_{k,2}(s_{k,1},m_k),\ldots,f_{k,n}(s_k^{n-1},m_k)
\bigr).
\]
The rate pair $(R_1,R_2)$ of the code is defined as
\[
R_1=\one\log \nu_1,\ \ \ \ R_2=\one\log \nu_2.
\]
A rate-pair $(R_1,R_2,)$ is said to be achievable if for every
$\epsilon>0$ and sufficiently large $n$ there exists an
$(n,2^{nR_1},2^{nR_2},\epsilon)$ code with strictly-causal SI for the
channel $P_{Y|S,X_1,X_2}$.  The capacity region of the channel
with strictly-causal independent SI is the closure of the set of all
achievable pairs $(R_1,R_2)$, and is denoted $\CapSCind$. The
superscript ``$\text{ind}$'' indicates that the two states are
independent.  %Our interest is in $\CapSCind$.

\section{The Single-State Scenario}
For the single-state scenario, an inner bound on $\CapSCcom$ was
derived in \cite{LapidothSteinbergIZS10} and later extended to
many-transmitters in \cite{LiSimeoneYenerArxiv}. In the absence of
cost constraints this bound can be described as follows:
Let $\pmfSCcom$ be the collection of all random variables 
$(U,V,X_1,X_2,\Scom,Y)$ whose joint distribution satisfies
\be
P_{U,V,X_1,X_2,\Scom,Y}=  \label{eq:3}
 P_{\Scom} P_{X_1|U} P_{X_2|U} P_{U} P_{V|\Scom} P_{Y|\Scom,X_1,X_2}.
\ee
Note that~\eqref{eq:3} implies the Markov relations
 $X_1\markov U\markov X_2$ and $V\markov \Scom\markov Y$,
and that the triplet $(X_1,U,X_2)$ is independent of $(V,\Scom)$.
 Let $\RSCcom$ be the convex hull of the collection of all 
 $(R_1,R_2)$ satisfying
\begin{subequations}
\label{eq:ski10}
\begin{IEEEeqnarray}{rCl}
R_1 &\leq& I(X_1;Y|X_2,U,V)\label{eq:4}\\
R_2 &\leq& I(X_2;Y|X_1,U,V)\label{eq:5}\\
R_1+R_2 &\leq& I(X_1,X_2;Y|U,V)\label{eq:6}\\
 R_1+R_2 &\leq&  I(X_1,X_2,V;Y) - I(V;\Scom)\label{eq:7}
%\Gamma_k &\geq& \E{\phi_k(X_k)},\ \ \ \ \ \ k=1,2\label{eq:7.1}
\end{IEEEeqnarray}
\end{subequations}
for some $(U,V,X_1,X_2,\Scom,Y)\in\pmfSCcom$. 
\begin{theorem}[\cite{LapidothSteinbergIZS10}]
\label{theo:sc1}
$\RSCcom\subseteq \CapSCcom$.
\end{theorem}
The achievability of this region is based on a Block-Markov scheme
where at Block~$\nu+1$ the transmitters send fresh private messages as
well as a common message that is used to send a compressed version of
the state sequence of Block~$\nu$. The compression is of the Wyner-Ziv
type with the side information being the channel outputs at
Block~$\nu$.

We next present a tighter inner bound. At Block~$\nu+1$ we still use
the MAC by sending private messages and a common message. The common
message is still a compressed version of the state information from
the previous block. The twist, however, is that the private messages
need not be entirely composed of fresh information. The private
message of Transmitter~1 has two parts. The first, of rate $R_{1}$, is
indeed fresh information. But the second, of rate $R_{0}^{(1)}$, is a
compressed version of the pair of sequences
$(\vect{x}_{1},\vect{\scom})$ from Block~$\nu$ (again with the side
information being the received symbols in the previous block). Since
Transmitter~1 knows which symbols it sent in the previous block, and
since it knows the state of the channel in the previous block, it can
compress the pair $(\vect{x}_{1},\vect{\scom})$. Likewise
Transmitter~2. Using Gastpar's results on the compression of
correlated sources with side information \cite{Gastpar2003} we obtain
the following bound:
\begin{theorem}
  \label{thm:amos345}
The rate-pair $(R_{1}, R_{2})$ is achievable if for some joint
distribution of the form
\begin{multline}
   \label{eq:newP3_11}
   P_{U,V,V_{1},V_{2},X_1,X_2,\Scom,Y}= \\
   P_{\Scom} P_{X_1|U} P_{X_2|U} P_{U} 
   P_{V|\Scom} P_{V_{1}|\Scom, X_{1}} P_{V_{2}|\Scom, X_{2}}
   P_{Y|\Scom,X_1,X_2}
\end{multline}
there exist nonnegative numbers $R_{0}^{(1)}$ and $R_{0}^{(2)}$ such
that
%\begin{subequations}
\begin{IEEEeqnarray}{rCl}
R_1 + R_{0}^{(1)} &\leq& I(X_1;Y, V_{1}, V_{2}, V|X_2,U)\label{eq:newP3_4}
\IEEEeqnarraynumspace \\
R_2 + R_{0}^{(2)} &\leq& I(X_2;Y, V_{1}, V_{2}, V|X_1,U)\label{eq:newP3_5}
\IEEEeqnarraynumspace \\
R_1+R_2 + R_{0}^{(1)} + R_{0}^{(2)} &\leq&
I(X_1,X_2;Y,V_{1}, V_{2}, V|U)\label{eq:newP3_6}
\IEEEeqnarraynumspace \\
R_0+R_1+R_2 + R_{0}^{(1)} + R_{0}^{(2)} &\leq&
I(X_1,X_2;Y,V_{1}, V_{2}, V)\label{eq:newP3_8} \IEEEeqnarraynumspace
\end{IEEEeqnarray}
%\end{subequations}
and
\begin{subequations}
\begin{IEEEeqnarray}{rCl}
  R_{0}^{(1)} & \geq & I(X_{1}, W; V_{1}|V,V_{2},Y) \label{eq:ligong81}\\
  R_{0}^{(2)} & \geq & I(X_{2}, W; V_{2}|V,V_{1},Y) \\
  R_{0} & \geq & I(W; V|V_{1},V_{2},Y) \\
  R_{0}^{(1)} + R_{0}^{(2)} & \geq & I(X_{1}, X_{2}, W; V_{1},
  V_{2}|V,Y) \\
  R_{0}^{(1)} + R_{0}& \geq & I(X_{1}, W; V_{1},
  V|V_{2},Y) \\
R_{0}^{(2)} + R_{0}& \geq & I(X_{2}, W; V_{2},
  V|V_{1},Y) \\
R_{0}^{(1)} + R_{0}^{(2)} + R_{0} & \geq & 
I(X_{1}, X_{2}, W; V_{1}, V_{2},V|Y). \label{eq:ligong87}
\end{IEEEeqnarray}
\end{subequations}
\end{theorem}
If we only consider joint distributions where $V_{1}$ and $V_{2}$ are
deterministic, and if we set $R_{0}^{(1)}$, $R_{0}^{(2)}$ to zero, we
obtain the inner bound of \cite{LapidothSteinbergIZS10}. Thus,
\begin{remark}
  The proposed inner bound contains the inner bound of
  \cite{LapidothSteinbergIZS10}
\end{remark}
The following example shows that the inclusion can be strict.
%The following example demonstrates that the inner bound $\RSCcom$ need
%not be tight, i.e., that for some channels the inclusion
%in Theorem~\ref{theo:sc1} is strict.
\begin{example}
  \label{ex:amos}
  Consider a MAC with two binary inputs $\set{X}_{1} = \set{X}_{2} =
  \{0,1\}$; a common state $\Scom = (\Scom_{0}, \Scom_{1}) \in
  \{0,1\}^{2}$, where $\Scom_{0}$, $\Scom_{1}$ are IID with entropy
  \begin{equation}
    \label{eq:pessach100}
    H(\Scom_{0}) = H(\Scom_{1}) = 1/2;
  \end{equation}
  and an output $Y = (Y_{1},Y_{2}) \in \{0,1\}^{2}$ with
  \begin{subequations}
    \label{eq:pessach110}
    \begin{align}
    Y_{1} & = X_{1} \oplus \Scom_{X_{2}} \\
    Y_{2} & = X_{2}.
  \end{align}
  \end{subequations}
  Thus, if $X_{2}$ is equal to zero, then $Y_{1}$ is the mod-2 sum of
  $X_{1}$ and $\Scom_{0}$, and otherwise it is the mod-2 sum of $X_{1}$ and
  $\Scom_{1}$. We study the highest rate at which User~2 can communicate
  when User~1 transmits at rate $1$. We show that for this channel
  \begin{equation}
    \label{eq:pessach120}
    \max \{ R_{2}: (1,R_{2}) \in \RSCcom \} = 0
  \end{equation}
  but
  \begin{equation}
    \label{eq:pessach130}
    \max \{ R_{2}: (1,R_{2}) \in \CapSCcom \} = 1/2,
  \end{equation}
  and that the rate-pair $(1,1/2)$ is in the new inner bound.
\end{example}
\begin{IEEEproof}
  We first prove~\eqref{eq:pessach130}. To this end we note that if
  $(1,R_{2})$ is achievable, then $R_{2}$ cannot exceed $1/2$. This
  can be shown using the full-cooperation outer-bound
  \cite{LapidothSteinbergIZS10},
  %(Proposition~\ref{prop:P2_outer1}),
  which implies that $(R_{1}, R_{2})$ can only be achievable if $R_{1}
  + R_{2} \leq 3/2$. Of more interest to us is the fact that the
  rate-pair $(1,1/2)$ is achievable. We demonstrate this using the
  new inner bound. Indeed, it is straightforward to verify that setting
  \begin{subequations}
  \begin{equation}
    R_{0}^{(1)} = R_{0} = 0, \quad R_{0}^{(2)} = 1/2,
  \end{equation}
  \begin{equation}
    V = V_{1} = 0, \quad V_{2} = W_{X_{2}},
  \end{equation}
  \begin{equation}
    U = 0,
  \end{equation}
  \begin{equation}
    X_{1}, X_{2} \sim \text{IID Bernoulli $1/2$},
  \end{equation}
  and
  \begin{equation}
    (R_{1}, R_{2}) = (1,1/2)
  \end{equation}
  satisfies all the required inequalities.
\end{subequations}
This choice corresponds to the following Block-Markov scheme:
%using a
%  Block-Markov scheme.  Interestingly, the implementation of this
%  scheme does not require that the state be known to Transmitter~1; it
%  suffices that it be known (strictly causally) to Transmitter~2.
%  \begin{quote}
%    \texttt{Yossi, note that it suffices that the state be known to
%      Transmitter 2. This makes this example also relevant for the case
%      of independent states.} 
%  \end{quote}
  In the Block-Markov scheme Transmitter~1 sends its data uncoded. At
  Block~$b+1$ Transmitter~2 sends $n$ bits, half of which are fresh
  data bits and half of which are used to describe the $n$-length
  sequence $\vect{\scom}_{\vect{x}_{2}}$ of the previous block. Note
  that Transmitter~2 does not describe the entire state sequence
  $\vect{w}$ of the previous block but only
  $\vect{\scom}_{\vect{x}_{2}}$. This latter sequence is known
  to Transmitter~2 at the beginning of Block~$b+1$ thanks to the
  strictly-causal state information and because it knows the sequence
  $\vect{x}_{2}$ it transmitted in the previous block. And $n/2$ bits
  suffice to describe this sequence because
  % , it can calculate $\Scom_{\vect{x}_{2}}$. 
  $\Scom_{X_{2}}$ is of entropy $1/2$.  
%  This scheme achieves the
%  rate-pair $(1,1/2)$ and completes the proof of
%  \eqref{eq:pessach130}.

  We now turn to proving \eqref{eq:pessach120}. We fix some
  distribution $P_{U,V,X_1,X_2,\Scom,Y}$ of the form \eqref{eq:3}, we
  assume that $(R_{1}=1,R_{2})$ satisfy 
  Inequalities~\eqref{eq:ski10}, and we then prove that $R_{2}$ must
  be zero. Since $R_{1} = 1$ and since $\set{X}_{1}$ is binary,
  Inequality~\eqref{eq:4} must hold with equality, and $X_{1}$ must be
  independent of $(X_{2}, U, V)$. By \eqref{eq:3}, this implies that
  \begin{subequations}
    \begin{equation}
      \label{eq:pessach500}
      \text{$X_{1}$ is independent of $(X_{2},U,V,\Scom)$.}
    \end{equation}
    From \eqref{eq:4} (that we know holds with equality) and the fact
    that $R_{1} = 1$ we also infer that
    \begin{align}
      1 & = H(Y|X_{2}, U, V) - H(Y|X_{1}, X_{2}, U, V) \nonumber \\
      & = H(Y_{1}|X_{2}, U, V) - H(Y_{1}|X_{1}, X_{2}, U, V)
      \label{eq:pessach505}
    \end{align}
    where the second equality holds because $Y_{2}$ is a
    deterministic function of $X_{2}$.  Since $Y_{1}$ is binary,
    $H(Y_{1}|X_{2}, U, V)$ is upper-bounded by $1$, and we
    conclude from \eqref{eq:pessach505} that
    \begin{align}
      0 & = H(Y_{1}|X_{1}, X_{2}, U, V) \nonumber \\
      & = H(Y_{1} \oplus X_{1}|X_{1}, X_{2}, U, V) \nonumber \\
      & = H(\Scom_{X_{2}}|X_{1}, X_{2}, U, V) \nonumber \\
      & = H(\Scom_{X_{2}}|X_{2}, U, V) 
      \label{eq:pessach510}      
    \end{align}
    where the last equality follows from
    \eqref{eq:pessach500}. 

    We next show that
    \begin{equation}
      \label{eq:pessach515}
      U \markov (X_{2}, V) \markov \Scom_{X_{2}}.
    \end{equation}
    To this end we note that, by \eqref{eq:3}, the pair $(V,\Scom)$ is
    independent of $(U,X_{2})$ and hence
    \begin{equation}
      \label{eq:pessach517}
      U \markov (X_{2},V) \markov \Scom.
    \end{equation}
    Since $\Scom_{X_{2}}$ is a deterministic function of
    $(X_{2},V, \Scom)$, this implies \eqref{eq:pessach515}, because
    if $A \markov B \markov C$ forms a Markov chain then 
    $A \markov B \markov f(B,C)$. Having established
    \eqref{eq:pessach515}, we now obtain from \eqref{eq:pessach510}
    \begin{equation}
      \label{eq:pessach519}
      H(\Scom_{X_{2}}|X_{2}, V) = 0.
    \end{equation}
    We now focus on the case where $X_{2}$ is not deterministic
    \begin{equation}
      \label{eq:pessach540}
      \bigPrv{X_{2} = \eta } > 0, \quad \eta \in \{0,1\},
    \end{equation}
    because if $X_{2}$ is deterministic then $R_{2}$ must be zero by
    \eqref{eq:5}. We also assume that the PMF of $V$ is strictly
    positive
    \begin{equation}
      \label{eq:pessach542}
      \bigPrv{V=v} > 0, \quad v \in \set{V},
    \end{equation}
    because outcomes of the auxiliary random variable that have zero
    probability can be removed from $\set{V}$ without affecting the
    inner bound. Since , by \eqref{eq:3}, $V$ is independent of
    $X_{2}$, it follows from \eqref{eq:pessach540} and
    \eqref{eq:pessach542} that
    \begin{equation}
      \label{eq:pessach530}
      \bigPrv{X_{2} = \eta, V = v} > 0, \quad \eta \in \{0,1\}, \; v
      \in \set{V}.
    \end{equation}
    This and \eqref{eq:pessach519} imply that
    \begin{equation}
      \label{eq:pessach600}
      H(\Scom_{\eta}|X_{2} = \eta, V = v) = 0, 
      \quad \eta \in \{0,1\}, \; v \in \set{V}.
    \end{equation}
    Since, by \eqref{eq:3}, $X_{2}$ is independent of $(V,\Scom)$
    and, \emph{a fortiori}, of $(V,\Scom_{\eta})$, it follows from
    \eqref{eq:pessach600} that 
    \begin{equation}
      \label{eq:pessach610}
      H(\Scom_{\eta}|V = v) = 0, 
      \quad \eta \in \{0,1\}, \; v \in \set{V}.
    \end{equation}
    Thus, $H(\Scom_{\eta}|V) = 0$, and since $\Scom = (\Scom_{0},
    \Scom_{1})$,
    \begin{equation}
      \label{eq:pessach620}
      H(\Scom|V) = 0.
    \end{equation}
    Consequently,
    \begin{align}
      I(V;\Scom) & = H(\Scom) \nonumber \\
      & = 1,
      \label{eq:pessach550}
    \end{align}
    where the second equality follows from~\eqref{eq:pessach100} and
    the independence of $\Scom_{0}$ and $\Scom_{1}$.  From
    \eqref{eq:pessach550}, \eqref{eq:7}, and the fact that $\set{Y}$
    has four elements we then conclude that $R_{1} + R_{2} \leq
    1$. This combines with $R_{1} = 1$ to establish that $R_{2}$ must
    be zero.
    \end{subequations}
\end{IEEEproof}

\noindent
\textbf{Terminating the Block-Markov scheme:} To conclude the sketch
of the achievability of the new inner bound, we still need to describe
how the Block-Markov scheme is terminated. We thus assume that $B$
blocks have been transmitted, and we proceed to describe Blocks $B+1$,
$B+2$, and $B+3$. We think about these blocks as ``overhead,'' because
they contain no fresh information. Fortunately, this overhead does not
affect the throughput because we can choose $B$ very large.

The next lemma shows that if the full-cooperation capacity of the MAC
without SI is zero, then the new inner bound contains only the rate-pair
$(0,0)$ and is thus trivially an inner bound.
\begin{lemma}
\label{lem:coop_zero}
If the capacity of the MAC without any side information but with full
cooperation is zero, i.e., if
\begin{equation}
  \label{eq:NoStateFullCooZero}
  \max_{P_{X_{1},X_{2}}} I(X_{1}, X_{2}; Y) = 0,
\end{equation}
then the proposed new inner bound contains only the all-zero rate
tuple.
\end{lemma}
\begin{IEEEproof}
  By \eqref{eq:ligong87} and \eqref{eq:newP3_8}, we conclude that if
  $R_{1},R_{2}$ is in the new inner bound, then for some joint
  distribution of the form~\eqref{eq:newP3_11}
  \begin{IEEEeqnarray*}{rCl}
    \IEEEeqnarraymulticol{3}{l}{R_{1} + R_{2} }\nonumber\\\quad
    & \leq &  I(X_1,X_2;Y,V_{1}, V_{2}, V) 
    - I(X_{1}, X_{2}, W; V_{1}, V_{2},V|Y) \\
    & = & I(X_1,X_2;Y) + I(X_1,X_2; V_{1}, V_{2}, V|Y) \nonumber \\ \quad
    & & - I(X_{1}, X_{2}, W; V_{1}, V_{2},V|Y).
  \end{IEEEeqnarray*}
  Consequently, if \eqref{eq:NoStateFullCooZero} holds and hence
  $I(X_1,X_2;Y)$ is zero, then $R_{1} + R_{2}$ must be upper-bounded
  by $I(X_1,X_2; V_{1}, V_{2}, V|Y) - I(X_{1}, X_{2}, W; V_{1},
  V_{2},V|Y)$, which is nonpositive.
\end{IEEEproof}

In view of Lemma~\ref{lem:coop_zero}, it only remains to prove the
achievability of the new inner bound when the full-cooperation
capacity without SI is positive. The next lemma shows that we can also
assume that the channel between Transmitter~1 (uninformed) and the
receiver (informed) is of positive capacity and likewise from
Transmitter~2.
\begin{lemma}
\label{lem:ToEachInformed}
If the channel between Transmitter~1 (uninformed) to the receiver
(informed) is of zero capacity, i.e., 
\begin{equation}
  \label{eq:InAir10}
  \max_{x_{2} \in \set{X}_{2}} \max_{P_{X_{1}}} I(X_{1};Y,W|X_{2} = x_{2})
  = 0,
\end{equation}
then the new inner bound contains only rate pairs $(R_{1},R_{2})$ with
$R_{1} = 0$ and $R_{2} \leq \max I(X_{2};Y)$. An analogous result
holds if
\begin{equation}
  \label{eq:InAir11}
  \max_{x_{1} \in \set{X}_{1}} \max_{P_{X_{2}}} I(X_{2};Y,W|X_{1} = x_{1})
  = 0,
\end{equation}

\end{lemma}
\begin{IEEEproof}
  We first prove that if a rate pair $(R_{1},R_{2})$ is in the new
  inner bound, and if \eqref{eq:InAir10} holds, then $R_{1}$ must be
  zero. Fix some joint distribution of the form~\eqref{eq:newP3_11}
  and let $(R_{1}, R_{2})$ satisfy the inequalities of
  Theorem~\ref{thm:amos345}.  We next argue that
  Hypothesis~\eqref{eq:InAir10} implies
  \begin{equation}
    \label{eq:Chianti10}
    I(X_1;Y, V_{2}, V|X_2,U) = 0.
  \end{equation}
Indeed,
\begin{subequations}
  \begin{IEEEeqnarray}{rCl}
    \IEEEeqnarraymulticol{3}{l}{I(X_1;Y, V_{2}, V|X_2,U) 
    }\nonumber\\\quad
    & \leq & I(X_1;Y, V_{2}, V|X_2,U,\Scom) \\
    & = & I(X_{1};Y|X_2,U,\Scom,V_{2}, V) \\
    & = & I(X_{1};Y|X_{2}, U, \Scom), \\
    & = & I(X_{1};Y, \Scom|X_{2}, U), \\
    & \leq & \max_{u \in \set{U}} \max_{x_{2} \in \set{X}_{2}} 
    I(X_{1};Y, \Scom|X_{2} = x_{2}, U = u) \\
    & \leq & \max_{u \in \set{U}} \max_{x_{2} \in \set{X}_{2}} 
    \max_{P_{X_{1}|U=u}} I(X_{1};Y, \Scom|X_{2} = x_{2}, U = u)
    \label{eq:sigh10} \IEEEeqnarraynumspace \\
    & = & \max_{x_{2} \in \set{X}_{2}} 
    \max_{P_{X_{1}}} I(X_{1};Y, \Scom|X_{2} = x_{2})
  \end{IEEEeqnarray}
\end{subequations}
where the first line follows from
\begin{equation}
  \label{eq:Mar10}
  X_{1} \markov (X_{2},U) \markov \Scom;
\end{equation}
the second from the chain rule and because 
\begin{equation}
  \label{eq:Mar20}
  X_{1} \markov (X_{2}, U, \Scom) \markov (V_{2},V)
\end{equation}
so $I(X_{1}; V_{2}, V|X_{2}, U, \Scom)$ is zero; the third
from
\begin{equation}
  \label{eq:Mar30}
  (X_{1},Y) \markov (X_{2}, U, \Scom) \markov (V_{2}, V);
\end{equation}
the fourth again by \eqref{eq:Mar10}; the fifth by upper bounding the
average by the maximal value; the sixth by maximizing over the
conditional distribution of $X_{1}$ given $U=u$; and the last because
the maximization over $u$ on the RHS of \eqref{eq:sigh10} is
unnecessary.

Continuing our proof that $R_{1}$ must be zero, we note
that~\eqref{eq:newP3_4} and \eqref{eq:ligong81} imply
\begin{IEEEeqnarray*}{rCl}
  R_{1} & \leq & I(X_1;Y, V_{1}, V_{2}, V|X_2,U) - I(X_{1}, W;
  V_{1}|V,V_{2},Y) \\
  & = & I(X_1;Y, V_{2}, V|X_2,U) + I(X_1;V_{1}| X_{2}, U, Y, V_{2}, V) 
  \nonumber \\
  && \quad -  I(X_{1}, W; V_{1}|V,V_{2},Y) \\
  & = & I(X_1;V_{1}| X_{2}, U, Y, V_{2}, V) - 
  I(X_{1}, W; V_{1}|V,V_{2},Y) \\
  & = & H(V_{1}| X_{2}, U, Y, V_{2}, V) - H(V_{1}| X_{1}, X_{2}, U, Y,
  V_{2}, V) \nonumber \\
  && \quad + H(V_{1}|X_{1}, W, V,V_{2},Y) -  H(V_{1}|V,V_{2},Y) \\
  & = & H(V_{1}| X_{2}, U, Y, V_{2}, V) - H(V_{1}| X_{1}, X_{2}, U, Y,
  V_{2}, V) \nonumber \\
  && \quad + H(V_{1}|X_{1}, W) -  H(V_{1}|V,V_{2},Y) \\
  &\leq & 0,
\end{IEEEeqnarray*}
where the second equality (third line) follows
from~\eqref{eq:Chianti10}, and where in the last inequality we have
used
\begin{equation*}
  H(V_{1}| X_{2}, U, Y, V_{2},V) \leq H(V_{1}|V,V_{2},Y)
\end{equation*}
(conditioning reduces entropy) and
\begin{equation*}
  H(V_{1}| X_{1}, X_{2}, U, Y, V_{2},V) \geq H(V_{1}|X_{1},W),
\end{equation*}
which can be argued as follows:
\begin{IEEEeqnarray*}{rCl}
  H(V_{1}| X_{1}, X_{2}, U, Y, V_{2},V) & \geq & H(V_{1}| X_{1}, W,
  X_{2}, U, Y, V_{2},V) \\ 
  & = & H(V_{1}| X_{1}, W),
\end{IEEEeqnarray*}
where the first inequality is because conditioning cannot increase
entropy, and the second by \eqref{eq:newP3_11}, which implies that,
conditional on $(X_{1},W)$, the auxiliary random variable $V_{1}$ is
independent of $(X_{2}, U, Y, V_{2},V)$.

Having established that $R_{1}$ is zero, we now conclude from
\eqref{eq:ligong87} and \eqref{eq:newP3_8}
\begin{IEEEeqnarray*}{rCl}
  R_{2} & = & R_{1} + R_{2} \\
  & \leq & I(X_1,X_2;Y,V_{1}, V_{2}, V) 
    - I(X_{1}, X_{2}, W; V_{1}, V_{2},V|Y) \\
  & = & I(X_{1},X_{2};Y) + \\
  & & \;
  I(X_{1},X_{2}; V_{1}, V_{2}, V|Y) - I(X_{1}, X_{2}, W; V_{1},
  V_{2},V|Y) \\
  & \leq & I(X_{1},X_{2};Y) \\
  & = & I(X_{2}; Y) + I(X_{1}; Y|X_{2}) \\
  & = & I(X_{2}; Y).
\end{IEEEeqnarray*}
\end{IEEEproof}

Lemma~\ref{lem:ToEachInformed} shows that if either \eqref{eq:InAir10}
or \eqref{eq:InAir11} holds, then the new inner bound is
achievable. It thus only remains to prove its achievability when 
\begin{equation}
  \label{eq:coffee50}
  \max_{x_{2} \in \set{X}_{2}} \max_{P_{X_{1}}} I(X_{1};Y,W|X_{2} = x_{2})
  > 0
\end{equation}
and
\begin{equation}
  \label{eq:coffee60}
  \max_{x_{1} \in \set{X}_{1}} \max_{P_{X_{2}}} I(X_{2};Y,W|X_{1} = x_{1})
  > 0,
\end{equation}
both of which we now assume.

We are now ready to describe the termination of the Block-Markov
scheme. Block~$B+1$ is split into two parts. In the first,
Transmitter~1 sends the $\vect{v}_{1}$-sequence of Block~$B$ assuming
that the receiver knows the state sequence $\vect{\scom}$ of
Block~$B+1$. This can be done (under this assumption) by \eqref{eq:coffee50}.
% because we have assumed that
% \begin{equation}
%   \max_{x_{2} \in \set{X}_{2}} \max_{P_{X_{1}}} I(X_{1};Y,W|X_{2} = x_{2})
%   > 0.
% \end{equation}
In the second, Transmitter~2 sends the $\vect{v}_{2}$-sequence of
Block~$B$ assuming that the receiver knows the state of
Block~$B+1$. This is possible by \eqref{eq:coffee60}.  
% because we have assumed that
% \begin{equation}
%   \max_{x_{1} \in \set{X}_{1}} \max_{P_{X_{2}}} I(X_{2};Y,W|X_{1} = x_{1})
%   > 0.
% \end{equation}
In Block~$B+2$ the transmitters cooperate to send the sequence
$\vect{\scom}$ of Block~$B+1$, and in Block~$B+3$ they
cooperate to send the $\vect{v}$ sequence of Block~$B$.

Decoding is performed as follows. The decoder first decodes
Block~$B+3$ without any side-information and thus learns the sequence
$\vect{v}$ of Block~$B$. It then decodes Block~$B+2$ (again without
any side information) and learns the state sequence~$\vect{\scom}$ of
Block~$B+1$. Now that it knows the state sequence of Block-$B+1$, it
can decode that block and learn the $\vect{v}_{1}$-sequence and the
$\vect{v}_{2}$-sequence of Block~$B$. From here on, it can proceed
with the regular backward decoding: in decoding Block~$b$ it knows the
sequences $\vect{v}$, $\vect{v}_{1}$, and $\vect{v}_{2}$ of Block~$b$
and it can therefore decode the common message and the messages
transmitted by each of the transmitters in Block~$b$. From this
decoding it learns the private messages of Block~$b$, and the
sequences $\vect{v}$, $\vect{v}_{1}$, and $\vect{v}_{2}$ of
Block~$b-1$.

\section{The Double-State Scenario}

For the double-state scenario, an inner bound on $\CapSCind$ was
proposed in \cite{LapidothSteinbergISIT10}. In the absence of cost
constraints this bound can be described as follows: Let $\pmfSCind$ be
the collection of all random variables $(V_1,V_2,S_1,S_2,X_1,X_2,Y)$
whose joint distribution satisfies
\begin{multline}
\label{eq:P2_4}
P_{V_1,V_2,S_1,S_2,X_1,X_2,Y}= \\
P_{V_1|S_1} P_{V_2|S_2} P_{S_1} P_{S_2} P_{X_1} P_{X_2} P_{Y|S_1,S_2,X_1,X_2}.
\end{multline}
Note that~\eqref{eq:P2_4} implies the Markov relations
 \begin{IEEEeqnarray}{l}
 V_1\markov S_1\markov (V_2,Y,S_2)\nonumber\\
 V_2\markov S_2\markov (V_1,Y,S_1)\nonumber\\
 (V_1,V_2)\markov(S_1,S_2)\markov Y \label{eq:P2_5}
 \end{IEEEeqnarray}
and that  $X_1,X_2$ are independent of each other and
of the quadruple $(V_1,V_2,S_1,S_2)$.
 Let $\RSCind$ be the convex hull of the collection of all 
 rate-pairs $(R_1,R_2)$ satisfying
\begin{IEEEeqnarray}{rCl}
0\leq R_1 &\leq& I(X_1;Y|X_2,V_1,V_2)-I(V_1;S_1|Y,V_2)\label{eq:P2_6_1}\\
0\leq R_2 &\leq& I(X_2;Y|X_1,V_1,V_2)-I(V_2;S_2|Y,V_1)\label{eq:P2_6_2}\\
R_1+R_2 &\leq&
I(X_1,X_2;Y|V_1,V_2)-I(V_1,V_2;S_1,S_2|Y)\label{eq:P2_6_3}
\IEEEeqnarraynumspace
%\\
%\Gamma_k &\geq& \E{\phi_k(X_k)},\ \ \ \ \ \ k=1,2\label{eq:P2_6_4}
\end{IEEEeqnarray}
for some $(V_1,V_2,S_1,S_2,X_1,X_2,Y)\in \pmfSCind$.
%Our main result for the strictly-causal case is the following.
\begin{theorem}[\cite{LapidothSteinbergISIT10}]
\label{theo:P2_sc1}
$\RSCind\subseteq \CapSCind$.
\end{theorem}
The proof is based on a scheme where lossy versions of the state
sequences are conveyed to the decoder using distributed Wyner-Ziv
compression~\cite{Gastpar2003} and Block-Markov encoding for the MAC,
to transmit the messages and the Wyner-Ziv codewords.  The channel
output serves as the decoder's SI in the distributed Wyner-Ziv code.
Since the two components of the source are independent, there is no
direct cooperation between the encoders via a common message as in
single-state scenario. Instead, each user spends part of its private
rate on the transmission of its Wyner-Ziv codeword.  

An improved inner bound was proposed by Li \emph{et al.} in
\cite{LiSimeoneYenerArxiv}. There it was shown that the improved inner
bound always contains the inner bound of
\cite{LapidothSteinbergISIT10}, and it was conjectured that there are
cases where the inclusion is strict. We next present the inner bound
of Li \emph{et al.} and then show that the inclusion can, indeed, be
strict.

Li \emph{et al.} consider all joint distributions of the form
\begin{multline}
  P_{V_{1}, V_{2}, S_{1}, S_{2}, X_{1}, X_{2},Y} = \\
  P_{V_{1}|S_{1},X_{1}} P_{V_{2}|S_{2},X_{2}} P_{S_{1}} P_{S_{2}}
  P_{X_{1}} P_{X_{2}} P_{Y|S_{1}, S_{2}, X_{1}, X_{2}}
\end{multline}
and prove the achievability of rate pairs $(R_{1}, R_{2})$ satisfying
\begin{subequations}
\begin{IEEEeqnarray}{rCl}
  R_{1} & \leq & I(X_{1},V_{1};Y|X_{2}, V_{2}) - I(V_{1}; S_{1}|X_{1}) \\
  R_{2} & \leq & I(X_{2},V_{2};Y|X_{1}, V_{1}) - I(V_{2}; S_{2}|X_{2}) \\
R_{1} + R_{2} & \leq  & I(X_{1}, X_{2}, V_{1}, V_{2}; Y) \nonumber \\ 
  && \quad - I(V_{1};S_{1}|X_{1}) - I(V_{2}; S_{2}|X_{2}). 
\end{IEEEeqnarray}
\end{subequations}
Roughly speaking, the improvement in the inner bound is the result of
Transmitter~1 compressing the pair $(\vect{s}_{1}, \vect{x}_{1})$ from
the previous block (with the outputs from the previous block serving
as side information) and not just $\vect{s}_{1}$ and likewise for
Transmitter~2. We next show, by example, that the bound of Li \emph{et
  al.} can, indeed, be tighter than that of Theorem~\ref{theo:P2_sc1}

The example is very similar to Example~\ref{ex:amos}. In fact, the
channel is as in Example~\ref{ex:amos}, but with the state 
$S_{1}$ being null (deterministic) and the state $S_{2}$ consisting of
the pair $(\Scom_{0}, \Scom_{1})$ of Example~\ref{ex:amos}:
\begin{subequations}
  \begin{equation}
    S_{1} = 0 \quad S_{2} = (\Scom_{0}, \Scom_{1}),
  \end{equation}
  where $\Scom_{0}, \Scom_{1}$ are IID binary random variables, each
  of entropy $1/2$.
%\end{subequations}

The rate pair $(R_{1}, R_{2}) = (1,1/2)$ is in the inner bound of Li
\emph{et al.}. To see this we set $V_{1} = 0$ and $V_{2} =
\Scom_{X_{2}}$ with $X_{1}$, $X_{2}$ IID random bits. However, as we
next prove, the pair $(1,1/2)$ is not in $\RSCind$. 

We prove this by showing that if $(1,R_{2})$ is in $\RSCind$,
then~$R_{2}$ must be zero. Suppose then that $(1,R_{2}) \in
\RSCind$. Since~$S_{1}$ is null, it follows from the structure
\eqref{eq:P2_4} of the joint distribution, that $V_{1}$ must be
independent of all the other random variables. Consequently, we can
strike it out from \eqref{eq:P2_6_1}, \eqref{eq:P2_6_2}, and
\eqref{eq:P2_6_3}. Since $R_{1} = 1$, it follows from
\eqref{eq:P2_6_1} that $X_{1}$ must be Bernoulli(1/2) and that
$H(X_{1}|X_{2}, V_{2}, Y)$ must be zero. This implies that
$H(W_{X_{2}} |X_{2}, V_{2}, Y)$ must also be zero (because $X_{1} =
Y_{1} \oplus W_{X_{2}}$). Consequently, $H(W_{X_{2}} |X_{2}, V_{2},
Y_{1})$ must also be zero (because $Y_{2} = X_{2}$). This implies that
%\begin{subequations}
\begin{equation}
\label{eq:exam10}
H(W_{X_{2}} |X_{2}, V_{2}) = 0
\end{equation}
because $X_{1}$ is Bernoulli(1/2) and independent of $(X_{2}, V_{2},
\Scom)$, so $Y_{1}$, which is equal to $X_{1} \oplus \Scom_{X_{2}}$,
must also be independent of $(X_{2}, V_{2},
\Scom)$. Equation~\eqref{eq:exam10} is reminiscent of
\eqref{eq:pessach519} (with $V_{2}$ replacing $V$).

As in Example~\ref{ex:amos}, we now distinguish between two cases
depending on whether $X_{2}$ is deterministic or not. If it is
deterministic, then the rate $R_{2}$ must be zero by
\eqref{eq:P2_6_2}. Consider now the case when it is not. In this case
$\Prv{X_{2} = \eta}$ is positive for all $\eta \in \{0,1\}$. Since
  $V_{2}$ is independent of $X_{2}$ (by \eqref{eq:P2_4}), and since
  without changing the inner bound we can assume that $\Prv{V_{2} =
    v_{2}}$ is positive for all $v_{2} \in \set{V}_{2}$, it follows
  that in this case
  \begin{equation}
    \Prv{X_{2} = \eta, \, V_{2} = v_{2}} > 0, \quad \eta \in \{0,1\},
    \; v _{2} \in \set{V}_{2}.
  \end{equation}
  This combines with \eqref{eq:exam10} to imply that
  \begin{equation}
    H(W_{\eta} |X_{2} = \eta, V_{2} = v_{2}) = 0, \quad \eta \in \{0,1\},
    \; v _{2} \in \set{V}_{2}.
  \end{equation}
%  for all $x_{2} \in \{0,1\}$ and $v_{2} \in \set{V}_{2}$. 
  This implies that
  \begin{equation}
    H(W_{\eta} |V_{2} = v_{2}) = 0, \quad \eta \in \{0,1\},
    \; v _{2} \in \set{V}_{2},
  \end{equation}
  because, by \eqref{eq:P2_4}, $X_{2}$ is independent of $(V_{2},
  S_{2})$ and hence \emph{a fortiori} of $(V_{2}, W_{\eta})$. Thus,
  $H(W_{\eta} |V_{2}) = 0$, and since $S_{2} = (W_{0}, W_{1})$,
  \begin{equation}
    H(S_{2}|V_{2}) = 0.
  \end{equation}
% Suppose now that it is notIf it is not, then \eqref{eq:exam10} implies that
% $W_{X_{2}}$ must be a deterministic function of $X_{2}$ and $V_{2}$
% and hence (\emph{cf.} \eqref{eq:pessach550})
  Consequently,
\begin{equation}
  \label{eq:exam20}
  I(V_{2};S_{2})  = H(S_{2}) = 1.
\end{equation}
This implies that also
\begin{equation}
  \label{eq:exam25}
  I(V_{2};S_{2}|Y) = 1,
\end{equation}
because $Y$ is independent of $(V_{2}, S_{2})$. It now follows from
\eqref{eq:exam25}, the fact that $V_{1}$ is deterministic, and from
\eqref{eq:P2_6_2} that $R_{2}$ must be zero.
\end{subequations}

\section{Summary} 

We have presented an improved inner bound on the capacity region of
the memoryless multiple-access channel that is controlled by an IID
state that is known strictly causally to the two encoders. This bound
contains the bound of \cite{LapidothSteinbergIZS10}, and we have
provided an example showing that the inclusion can be strict.

We also adapted this example to a memoryless multiple-access channel
that is governed by two independent states, where each transmitter
knows one of the states strictly causally. The resulting example
demonstrates that---as conjecture by Li \emph{et al.}
\cite{LiSimeoneYenerArxiv}---the inner bound of Li \emph{et al.} can
be strictly tighter than that of \cite{LapidothSteinbergISIT10}.

\end{document}